# Nonparametric Bayes Pachinko Allocation

**Wei Li**
Department of Computer Science
University of Massachusetts
Amherst, MA 01003

**David Blei**
Computer Science Department
Princeton University
Princeton, NJ 08540

**Andrew McCallum**
Department of Computer Science
University of Massachusetts
Amherst, MA 01003

## Abstract

Recent advances in topic models have explored complicated structured distributions to represent topic correlation. For example, the pachinko allocation model (PAM) captures arbitrary, nested, and possibly sparse correlations between topics using a directed acyclic graph (DAG). While PAM provides more flexibility and greater expressive power than previous models like latent Dirichlet allocation (LDA), it is also more difficult to determine the appropriate topic structure for a specific dataset. In this paper, we propose a nonparametric Bayesian prior for PAM based on a variant of the hierarchical Dirichlet process (HDP). Although the HDP can capture topic correlations defined by nested data structure, it does not automatically discover such correlations from unstructured data. By assuming an HDP-based prior for PAM, we are able to learn both the number of topics and how the topics are correlated. We evaluate our model on synthetic and real-world text datasets, and show that nonparametric PAM achieves performance matching the best of PAM without manually tuning the number of topics.

## 1 Introduction

Statistical topic models such as latent Dirichlet allocation (LDA) (Blei et al., 2003) have been shown to be effective tools in topic extraction and analysis. These models can capture word correlations in a collection of textual documents with a low-dimensional set of multinomial distributions. Recent work in this area has investigated richer structures to also describe inter-topic correlations, and led to discovery of large numbers of more accurate, fine-grained topics.

One example is the correlated topic model (CTM) (Blei & Lafferty, 2006). CTM represents each document as a mixture of topics, where the mixture proportion is sampled from a logistic normal distribution. The parameters of this prior include a covariance matrix in which each entry specifies the covariance between a pair of topics. Therefore, topic occurrences are no longer independent from each other. However, CTM is limited to pairwise correlations only, and the number of parameters in the covariance matrix grows as the square of the number of topics.

An alternative model that provides more flexibility to describe correlations among topics is the pachinko allocation model (PAM) (Li & McCallum, 2006), which uses a directed acyclic graph (DAG) structure to represent and learn arbitrary-arity, nested, and possibly sparse topic correlations. Each leaf node in the DAG represents a word in the vocabulary, and each interior node corresponds to a topic. A topic in PAM can be not only a distribution over words, but also a distribution over other topics. Therefore it captures inter-topic correlations as well as word correlations. The distribution of a topic over its children could be parameterized arbitrarily. One example is to use a Dirichlet distribution, from which a multinomial distribution over its children is sampled on a per-document-basis. To generate a word in a document in PAM, we start from the root, samples a topic path based on these multinomials and finally samples the word.

The DAG structure in PAM is extremely flexible. It could be a simple tree (hierarchy), or an arbitrary DAG, with cross-connected edges, and edges skipping levels. Some other models can be viewed as special cases of PAM. For example, LDA corresponds to a three-level hierarchy consisting of one root at the top, a set of topics in the middle and a word vocabulary at the bottom. The root is fully connected to all the topics, and each topic is fully connected to all the words. The structure of CTM can be described with a four-level DAG in PAM, in which there are two levels of topic nodes between the root and words. The lower-level consists of traditional LDA-style topics. In the upper level there is one node for every topic pair, capturing pairwise correlations.

While PAM provides a powerful means to describe inter-topic correlations and extract large numbers of fine-grained topics, it has the same practical difficulty as many other topic models, *i.e.* how to determine the number of topics. It can be estimated using cross-validation, but this method is not efficient even for simple topic models like LDA. Since PAM has a more complex topic structure, it is more difficult to evaluate all possibilities and select the best one.

Another approach to this problem is to automatically learn the number of topics with a nonparametric prior such as the hierarchical Dirichlet process (HDP) (Teh et al., 2005). HDPs are intended to model groups of data that have a pre-defined hierarchical structure. Each pre-defined group is associated with a Dirichlet process whose base measure is sampled from a higher-level Dirichlet process. Note that HDP cannot automatically discover topic correlations from unstructured data. However, it has been applied to LDA where it integrates over (or alternatively selects) the appropriate number of topics.

In this paper, we propose a nonparametric Bayesian prior for pachinko allocation based on a variant of the hierarchical Dirichlet process. We assume that the topics in PAM are organized into multiple levels and each level is modeled with an HDP to capture uncertainty in the number of topics. Unlike a standard HDP mixture model, where the data has a pre-defined nested structure, we build HDPs based on dynamic groupings of data according to topic assignments. The nonparametric PAM can be viewed as an extension to fixed-structure PAM where the number of topics at each level is taken to infinity. To generate a document, we first sample multinomial distributions over topics from corresponding HDPs. Then we repeatedly sample a topic path according to the multinomials for each word in the document.

The rest of the paper is organized as follows. We detail the nonparametric pachinko allocation model in Section 2, describing its generative process and inference algorithm. Section 3 presents experimental results on synthetic data and real-world text data. We compare discovered topic structures with true structures for various synthetic settings, and likelihood on held-out test data with PAM, HDP and hierarchical LDA (Blei et al., 2004) on two datasets. Section 4 reviews related work, followed by conclusions and future work in Section 5.

## 2 The Model

### 2.1 Four-Level PAM

Pachinko allocation captures topic correlations with a directed acyclic graph (DAG), where each leaf node is associated with a word and each interior node corresponds to a topic, having a distribution over its children. An interior node whose children are all leaves would correspond to a traditional LDA topic. But some interior nodes may also have children that are other topics, thus representing a mixture over topics.

While PAM allows arbitrary DAGs to model topic correlations, in this paper we focus on one special class of structures. Consider a four-level hierarchy consisting of one root topic, $s_2$ topics at the second level, $s_3$ topics at the third level, and words at the bottom. We call the topics at the second level super-topics and the ones at the third level sub-topics. The root is connected to all super-topics, super-topics are fully connected to sub-topics and sub-topics are fully connected to words. For the root and each super-topic, we assume a Dirichlet distribution parameterized by a vector with the same dimension as the number of children. Each sub-topic is associated with a multinomial distribution over words, sampled once for the whole corpus from a single Dirichlet distribution. To generate a document, we first draw a set of multinomials from the Dirichlet distributions at the root and super-topics. Then for each word in the document, we sample a topic path consisting of the root, super-topic and sub-topic based on these multinomials. Finally, we sample a word from the sub-topic.

As with many other topic models, we need to specify the number of topics in advance for the four-level PAM. It is inefficient to manually examine every $(s_2, s_3)$ pair in order to find the appropriate structure. Therefore, we develop a nonparametric Bayesian prior based on the Dirichlet process to automatically determine the numbers of topics. As a side effect, this also discovers a sparse connectivity between super-topics and sub-topics. We present our model in terms of Chinese restaurant process.

### 2.2 Chinese Restaurant Process

The Chinese restaurant process (CRP) is a distribution on partitions of integers. It assumes a Chinese restaurant with an infinite number of tables. When a customer comes, he sits at a table with the following probabilities:

- $P(\text{an occupied table } t) = \frac{C(t)}{\sum_{t'} C(t') + \alpha}$,
- $P(\text{an unoccupied table}) = \frac{\alpha}{\sum_{t'} C(t') + \alpha}$,

| Name | Description |
|---|---|
| $r_j$ | the $j$th restaurant. |
| $e_{jk}$ | the $k$th entryway in the $j$th restaurant. |
| $c_l$ | the $l$th category. Each entryway $e_{jk}$ is associated with a category. |
| $t_{jln}$ | the $n$th table in the $j$th restaurant that has category $c_l$. |
| $m_{lp}$ | the $p$th menu in category $c_l$. Each table is associated with a menu from the corresponding category. |
| $d_m$ | the $m$th dish in the global set of dishes. Each menu is associated with a dish. |

Table 1: Notation for the generative process of nonparametric PAM.

where $C(t)$ is the number of customers sitting at table $t$, $\sum_{t'} C(t')$ is the total number of customers in the restaurant and $\alpha$ is a parameter. For the rest of the paper, we denote this process with $\mathrm{CRP}(\{C(t)\}_t, \alpha)$.

Note that the number of occupied tables grows as new customers arrive. After all customers sit down, we obtain a partition of integers, which exhibits the same clustering structure as draws from a Dirichlet process (DP) (Ferguson, 1973).

### 2.3 Nonparametric PAM

Now we describe the nonparametric prior for PAM as a variant of Chinese restaurant process.

#### 2.3.1 Generative process

Consider a scenario in which we have multiple restaurants. Each restaurant has an infinite number of *entryways*, each of which has a *category* associated with it. The restaurant is thus divided into an infinite number of sections according to the categories. When a customer enters an entryway, the category leads him to a particular section, where there are an infinity number of *tables*. Each table has a *menu* on it and each menu is associated with a *dish*. The menus are category-specific, but shared among different restaurants. Dishes are globally shared by all menus. If a customer sits at a table that already has other customers, he shares the same menu and thus dish with these customers. When he wants to sit at a new table, a menu is assigned to the table. If the menu is new too, a dish will be assigned to it.

A more formal description is presented below, using notation from Table 1.

A customer $x$ arrives at restaurant $r_j$.

1. He chooses the $k$th entryway $e_{jk}$ in the restaurant from $\mathrm{CRP}(\{C(j,k)\}_k, \alpha_0)$, where $C(j,k)$ is the number of customers that entered the $k$th entryway before in this restaurant.

2. If $e_{jk}$ is a new entryway, a category $c_l$ is assigned to it from $\mathrm{CRP}(\{\sum_{j'} C(l,j')\}_l, \gamma_0)$, where $C(l,j')$ is the number of entryways that have category $c_l$ in restaurant $r_{j'}$ and $\sum_{j'} C(l,j')$ is the total number of entryways in all restaurants that have category $c_l$.

3. After choosing the category, the customer makes the decision for which table he will sit at. He chooses table $t_{jln}$ from $\mathrm{CRP}(\{C(j,l,n)\}_n, \alpha_1)$, where $C(j,l,n)$ is the number of customers sitting at table $t_{jln}$.

4. If the customer sits at an existing table, he will share the menu and dish with other customers at the same table. Otherwise, he will choose a menu $m_{lp}$ for the new table from $\mathrm{CRP}(\{\sum_{j'} C(j',l,p)\}_p, \gamma_1)$, where $C(j',l,p)$ is the number of tables in restaurant $j'$ that have menu $m_{lp}$ and $\sum_{j'} C(j',l,p)$ is the number of tables in all restaurants that have menu $m_{lp}$.

5. If the customer gets an existing menu, he will eat the dish on the menu. Otherwise, he samples dish $d_m$ for the new menu from $\mathrm{CRP}(\{\sum_{l'} C(l',m)\}_m, \phi_1)$, where $C(l',m)$ is the number of menus in category $c_{l'}$ that serve dish $d_m$ and $\sum_{l'} C(l',m)$ is the number of all menus that serve dish $d_m$.

$\alpha_0$, $\alpha_1$, $\gamma_0$, $\gamma_1$ and $\phi_1$ are parameters in the Chinese restaurant processes.

Now we briefly explain how to generate a corpus of documents from PAM with this process. Each document corresponds to a restaurant and each word is associated with a customer. In the four-level DAG structure of PAM, there are an infinite number of super-topics represented as categories and an infinite number of sub-topics represented as dishes. Both super-topics and sub-topics are globally shared among all documents. To generate a word in a document, we first sample a super-topic according to the CRPs that sample the category. Then given the super-topic, we sample a sub-topic in the same way that a dish is drawn from CRPs. Finally we sample the word from the sub-topic according to its multinomial distribution over words. An example to generate a piece of text is shown in Figure 1.

As we can see, the sampling procedure for super-topics involves two levels of CRPs that draw entryways and categories respectively. They act as a prior on the distributions over super-topics, which is an example of hierarchical Dirichlet process (HDP). As Dirichlet processes have been used as priors for mixture models, the HDP can be used for problems where mixture components need to be shared among multiple DPs. One application of HDP is to automatically determine the number of topics in LDA, where each document is associated with one DP and topics are shared among all documents. Similarly, the prior on the distributions over sub-topics is also an HDP. It consists of three levels of CRPs to sample tables, menus and dishes. The hierarchy reflects groupings of the data. In addition to documents, the data is dynamically organized into groups based on super-topic assignments.

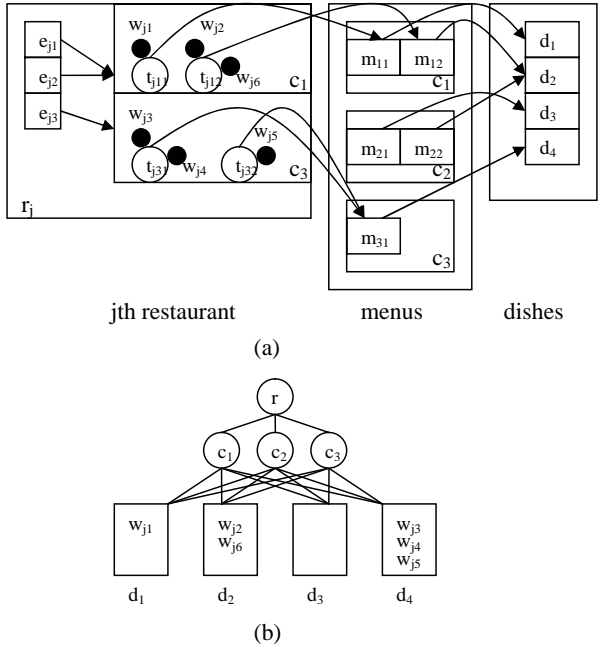

Figure 1: An example to demonstrate the generative process of nonparametric PAM. (a) shows the sampling result for a piece of text "$w_{j1}, ..., w_{j6}$" in the $j$th document. An arrow between two objects represents their association. For example, the category associated with entryways $e_{j1}$ and $e_{j2}$ is $c_1$, the menu associated with table $t_{j11}$ is $m_{11}$ and the dish associated with $m_{11}$ is $d_1$. (b) is the corresponding topic structure. The six words are generated from three different sub-topics.

### 2.3.2 Inference

Similar to PAM and many other topic models, we use Gibbs sampling to perform inference. For each customer $x$, we want to jointly sample the 5 variables associated with it. Assume $x$ wants to sit at restaurant $r_j$, then the conditional probability that he chooses entryway $e_{jk}$, category $c_l$, table $t_{jln}$, menu $m_{lp}$ and dish $d_m$ given observations $X$ and other variable assignments (denoted by $\Pi_{-x}$) is proportional to the product of the following terms:

1. $P(e_{jk}, c_l | \Pi_{-x}) \propto$

$$\begin{cases} \frac{C(j,k)}{\sum_{k'} C(j,k') + \alpha_0}, \\ \text{if } e_{jk} \text{ is an existing entryway with category } c_l; \\ \frac{\alpha_0}{\sum_{k'} C(j,k') + \alpha_0} \frac{\sum_{j'} C(l,j')}{\sum_{j'} \sum_{l'} C(l',j') + \gamma_0}, \\ \text{if } e_{jk} \text{ is new and } c_l \text{ is an existing category}; \\ \frac{\alpha_0}{\sum_{k'} C(j,k') + \alpha_0} \frac{\gamma_0}{\sum_{j'} \sum_{l'} C(l',j') + \gamma_0}, \\ \text{if } e_{jk} \text{ and } c_l \text{ are both new}; \\ 0, \text{other cases} \end{cases}$$

Note that the number of non-zero probabilities here is only the sum of the numbers of existing entryways and categories instead of their product.

2. $P(t_{jln}, m_{lp}, d_m | \Pi_{-x}, c_l) \propto$

$$\begin{cases} \frac{C(j,l,n)}{\sum_{n'} C(j,l,n') + \alpha_1}, \\ \text{if } t_{jln} \text{ is an existing table with } m_{lp} \text{ and } d_m; \\ \frac{\alpha_1}{\sum_{n'} C(j,l,n') + \alpha_1} \frac{\sum_{j'} C(j',l,p)}{\sum_{j'} \sum_{p'} C(j',l,p') + \gamma_1}, \\ \text{if } t_{jln} \text{ is new, but } m_{lp} \text{ is an existing menu with } d_m; \\ \frac{\alpha_1}{\sum_{n'} C(j,l,n') + \alpha_1} \frac{\gamma_1}{\sum_{j'} \sum_{p'} C(j',l,p') + \gamma_1} \frac{\sum_{l'} C(l',m)}{\sum_m \sum_{l'} C(l',m) + \phi_1}, \\ \text{if } t_{jln} \text{ and } m_{lp} \text{ are new and } d_m \text{ is an existing dish}; \\ \frac{\alpha_1}{\sum_{n'} C(j,l,n') + \alpha_1} \frac{\gamma_1}{\sum_{j'} \sum_{p'} C(j',l,p') + \gamma_1} \frac{\phi_1}{\sum_m \sum_{l'} C(l',m) + \phi_1}, \\ \text{if all three variables are new}; \\ 0, \text{other cases} \end{cases}$$

Again, the number of non-zero probabilities is not the product of the numbers of possible values for the three variables.

3. $P(x | X_{-x}, \Pi_{-x}, d_m) \propto \frac{C(m,x) + \beta}{\sum_{x'} (C(m,x') + \beta)}$

This calculates the probability to sample an observation $x$. Here we assume that the base distribution $H$ is a symmetric Dirichlet distribution with parameter $\beta$. $C(m, x)$ is the number of times that menu $d_m$ is assigned to customer $x$.

In order to use the above equations for Gibbs sampling, we still need to determine the values for the Dirichlet process parameters. Following the same procedure described in (Teh et al., 2005), we assume Gamma priors for $\alpha_0$, $\gamma_0$, $\alpha_1$, $\gamma_1$ and $\phi_1$ and re-sample them at every iteration.

## 3 Experimental Results

In this section, we describe experimental results for nonparametric PAM and several related topic models.

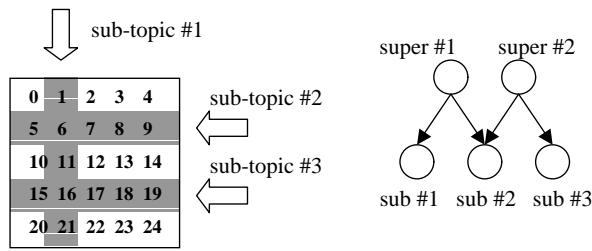

Figure 2: An example of synthetic topic structure

### 3.1 Datasets

#### 3.1.1 Synthetic datasets

One goal of nonparametric PAM is to automatically discover the topic structure from data. To demonstrate this ability for our model, we first apply it to synthetic datasets where the true structures are known. The word vocabulary of the training data is organized into a $v$-by-$v$ grid. Each sub-topic is a uniform distribution over either a column or a row of words, and each super-topic is an arbitrary combination of sub-topics. An example structure is shown in Figure 2. We follow the generative process of PAM with a fixed structure and Dirichlet parameters to randomly sample documents. By changing the vocabulary size, numbers of super-topics and sub-topics, we obtain different datasets where each of them consists of 100 documents and each document contains 200 tokens.

#### 3.1.2 20 newsgroups comp5 dataset

The 20 newsgroups dataset consists of 20,000 postings collected from 20 different newsgroups. In our experiment, we use the comp5 subset that includes 5 groups: graphics, os.ms-windows.misc, sys.ibm.pc.hardware, sys.mac.hardware and windows.x. There are in total 4,836 documents, 468,252 word tokens and 35,567 unique words. This dataset is specifically chosen because of the observed partitioning of documents into groups, which allows us to compare against HDP.

#### 3.1.3 Rexa dataset

*Rexa* is a digital library of computer science, which allows searching for research papers, authors, grants, etc (http://rexa.info). We randomly choose a subset of abstracts from its large collection. In this dataset, there are 5,000 documents, 350,760 word tokens and 25,597 unique words.

#### 3.1.4 NIPS dataset

This dataset consists of 1,647 abstracts of NIPS conference papers from 1987 to 1999. The vocabulary contains 11,708 words and the whole corpus has 114,142 word tokens in total.

| Model | Parameters |
|---|---|
| PAM | $\beta = 0.01$ |
| | $\alpha_r = 0.01$ |
| hLDA | $\beta = 0.01$ |
| | $\alpha = 10.0$ |
| | $\gamma = 1.0$ |
| HDP | $\beta = 0.01$ |
| | $\alpha \sim \text{Gamma}(1.0, 1.0)$ |
| | $\gamma \sim \text{Gamma}(1.0, 1.0)$ |
| | $\phi \sim \text{Gamma}(1.0, 10.0)$ |
| Nonparametric PAM | $\beta = 0.01$ |
| | $\alpha_0 \sim \text{Gamma}(1.0, 0.1)$ |
| | $\gamma_0 \sim \text{Gamma}(1.0, 1.0)$ |
| | $\alpha_1 \sim \text{Gamma}(1.0, 1.0)$ |
| | $\gamma_1 \sim \text{Gamma}(1.0, 1.0)$ |
| | $\phi_1 \sim \text{Gamma}(1.0, 10.0)$ |

Table 2: Model parameters: $\beta$ parameterizes the Dirichlet prior for topic distributions over words; $\alpha_r$ in PAM is the Dirichlet distribution associated with the root; in hLDA, $\alpha$ is the Dirichlet prior for distributions over topics and $\gamma$ is the parameter in the Dirichlet process; in HDP and nonparametric PAM, the parameters in Dirichlet processes are assumed to be drawn from various Gamma distributions.

### 3.2 Model Parameters

We assume the same topic structure for PAM, hLDA and nonparametric PAM except that hLDA and nonparametric PAM need to learn the numbers of topics. The parameters we use in the models are summarized in Table 2. In the training procedure, each Gibbs sampler is initialized randomly and runs for 1000 burn-in iterations. We then draw a total number of 10 samples with 100 iterations apart. The evaluation result is based on the average of these samples.

### 3.3 Discovering Topic Structures

We first apply nonparametric PAM to the synthetic datasets. After training, we associate each true topic with one discovered topic based on similarity and evaluate the performance based on topic assignment accuracy. The results are presented in Table 3. In all four experiments, sub-topics are identified with high accuracy. We have slightly lower accuracy on super-topics. The most common error is splitting. For example, in the second experiment (a 5-by-5 vocabulary, 3 super-topics and 7 sub-topics), one super-topic is split into three, which leads to a small decrease in accuracy.

| Structure | | | Accuracy (%) | |
|---|---|---|---|---|
| $v$ | $s_2$ | $s_3$ | Super-topic | Sub-topic |
| 5 | 2 | 4 | 99.93 | 98.78 |
| 5 | 3 | 7 | 82.03 | 97.70 |
| 10 | 3 | 7 | 99.43 | 95.86 |
| 10 | 4 | 10 | 96.14 | 96.66 |

Table 3: Evaluation result on synthetic datasets. Each row corresponds to one experiment. $v$ is the size of the grid, $s_2$ is the true number of super-topics and $s_3$ is the true number of sub-topics. Accuracy of super-topic (sub-topic) is the percentage of words whose super-topic (sub-topic) assignments are the same as the true topics.

### 3.4 Topic Examples

In addition to synthetic datasets, we also apply our model to real-world text data. For the 20 newsgroups comp5 dataset, we present example topics discovered by nonparametric PAM and PAM with 5 sub-topics (Table 4). As we can see, the 5 topics from the fixed-structure PAM are noisy and do not have clear correspondence to the 5 document groups. On the other hand, the topics from the nonparametric PAM are more salient and demonstrate stronger correlations with document groups. For example, the second topic is associated with documents in the sys.mac.hardware group and the fourth topic is associated with the graphics group. The comparison here shows the importance of choosing the appropriate number of topics.

We also present topic examples from the NIPS dataset in Figure 3. The circles ($\{t_1, ..., t_5\}$) correspond to 5 super-topics from 43 in total and each box displays the top 5 words in a sub-topic. A line between two topics represents their connection and its width approximately shows the proportion of a child topic in its parent. For example, in super-topic $t_3$, the dominant sub-topic is *neural networks*. But it also consists of other sub-topics such as *reinforcement learning* and stop words. Super-topics $t_1$ and $t_2$ share the same sub-topic about *classification*, which is discussed in two different contexts of *image recognition* and *neural networks*. Note that the discovered connectivity structure is sparse. For example, there is no connection between $t_4$ and the sub-topic about *image recognition*. It means that this sub-topic is never sampled from this super-topic in the whole dataset. The average number of children for a super-topic is 19 while the total number of sub-topics is 143.

| | 20ng | Rexa |
|---|---|---|
| PAM 5-100 | -797350 | |
| PAM 5-200 | -795760 | |
| PAM 20-100 | -792130 | -580373 |
| PAM 20-200 | -785470* | **-574964** |
| PAM 50-100 | -789740 | -577450 |
| PAM 50-200 | -784540* | -577086 |
| HDP | -791120 | |
| hLDA | -790312 | -581317 |
| NPB PAM | **-783298** | -575466* |

Table 5: Average log-likelihoods from 5-fold cross validation on 20 newsgroups comp5 dataset and Rexa dataset. The bold numbers are the highest values, and the ones with * are statistically equivalent to them.

### 3.5 Likelihood Comparison

We compare nonparametric PAM with other topic models in terms of likelihood on held-out test data. In order to calculate the likelihood, we need to integrate out the sampled distributions and sum over all possible topic assignments. This problem has no closed-form solution. Instead of approximating the likelihood of a document $d$ by the harmonic mean of a set of conditional probabilities $P(d|\mathbf{z}^{(d)})$ where the samples are generated using Gibbs sampling (Griffiths & Steyvers, 2004), we choose a more robust approach based on empirical likelihood (Diggle & Gratton, 1984). For each trained model, we first randomly sample 1,000 documents according to its own generative process. Then from each sample we estimate a multinomial distribution. The probability of a test document is then calculated as its average probability from each multinomial, just as in a simple mixture model.

We report evaluation results on two datasets. For the 20 newsgroups comp5 dataset, only HDP (Teh et al., 2005) uses the document group information in the training procedure. PAM, hLDA and nonparametric PAM do not rely on the pre-defined data structure. We use {5, 20, 50} super-topics and {100, 200} sub-topics for PAM, where the choice of 5 super-topics is to match the number of document groups. For the Rexa dataset, we do not include HDP since the data does not have a hierarchical structure comparable to other models. Thus there is no need to include PAM 5-100 and 5-200, which certainly would perform worse than other settings anyway. We conduct 5-fold cross validation and the average log-likelihood on test data is presented in Table 5.

Each row in the table shows the log-likelihoods of one model on the two datasets. The first column is the model name. For example, PAM 5-100 corresponds to PAM with 5 super-topics and 100 sub-topics, and the last three rows are HDP, hLDA and nonparametric PAM respectively.

| NPB PAM (179 sub-topics) | | | | | Fixed-structure PAM (5 sub-topics) | | | | |
|---|---|---|---|---|---|---|---|---|---|
| drive | mac | mb | jpeg | power | drive | file | windows | graphics | window |
| disk | system | simms | image | cpu | card | entry | file | image | server |
| drives | comp | ram | gif | fan | scsi | output | files | mail | motif |
| hard | disk | memory | color | heat | writes | program | jpeg | ftp | sun |
| controller | sys | bit | images | supply | mb | build | dos | data | widget |
| bios | ftp | vram | format | motherboard | system | line | don | software | application |
| floppy | macintosh | simm | quality | sink | article | entries | image | pub | display |
| system | apple | board | file | case | mac | printf | program | information | set |
| ide | faq | meg | bit | switch | problem | echo | writes | tax | mit |
| scsi | software | chip | version | chip | don | char | bit | package | xterm |

Table 4: Example topics in 20 newsgroups comp5 dataset. The left side are topics from nonparametric PAM and the right side are topics from PAM with 5 sub-topics. Each column displays the top 10 words in one sub-topic. By automatically choosing the number of topics, nonparametric PAM generates topics with higher quality than PAM with an inappropriate number of topics.

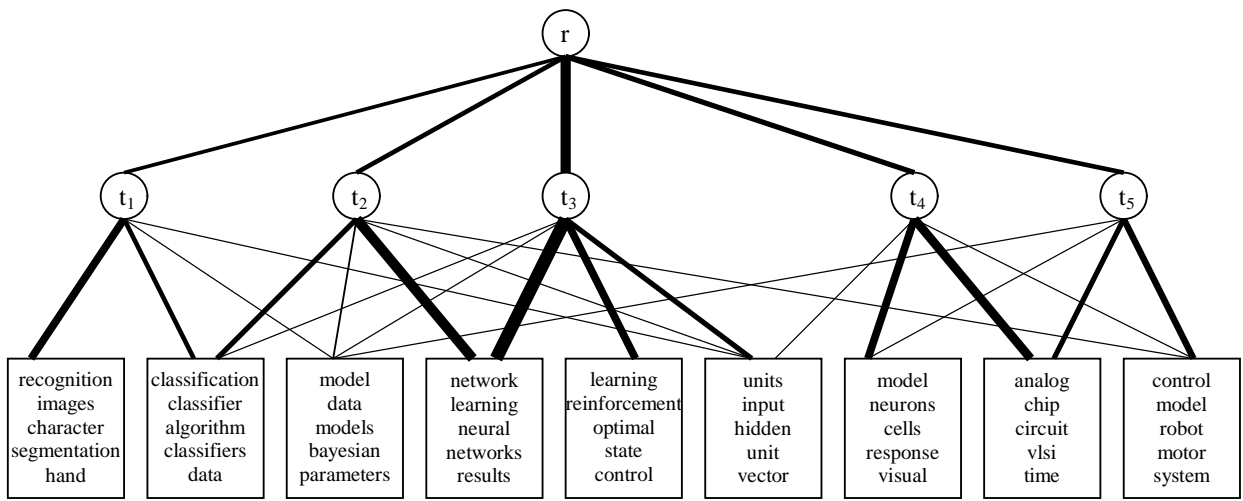

Figure 3: Example topics discovered by nonparametric PAM from NIPS dataset. Circles correspond to the root and super-topics, while each box displays the top 5 words in a sub-topic. The line width approximately shows the proportion of a child topic in its parent.

For the 20 newsgroups comp5 dataset, PAM does not perform as well as HDP when using only 5 super-topics. One possible reason is that HDP is provided with additional information of document groups while PAM is not. However, as more super-topics are used, PAM benefits from the ability of discovering more specific topics and eventually outperforms HDP. On the other hand, nonparametric PAM automatically discovers the topic structure from data. Its average number of super-topics is 67 and number of sub-topics is 173. Based on paired t-test results, nonparametric PAM performs as well as the best settings of PAM (20-200 and 50-200), and significantly better than both HDP and hLDA.

We obtain similar results for the Rexa dataset. The best manually-selected setting for PAM is 20 super-topics and 200 sub-topics. The nonparametric PAM performs just as well with an average number of 33 super-topics and 181 sub-topics. Again, hLDA is significantly worse according to paired t-tests.

## 4 Related Work

Choosing an appropriate number of mixture components is always an important issue for mixture models. Model selection methods such as cross-validation and Bayesian model testing are usually inefficient. A nonparametric solution with the Dirichlet process is more desirable because it does not require specifying the number of mixture components in advance. Dirichlet process mixture models have been widely studied in many problems (Kim et al., 2006; Daume-III & Marcu, 2005; Xing et al., 2004; Sudderth et al., 2005).

In order to solve problems where a set of mixture models share the same mixture components, Teh et al. (2005) propose the hierarchical Dirichlet process (HDP). One example of using the HDP is to learn the number of topics in LDA, in which each document is associated with a Dirichlet process whose base measure is sampled from a higher level Dirichlet process. Although it does not directly discover topic correlations from unstructured data, the HDP can be used as a nonparametric prior for other topic models such as PAM to automatically learn topic structures.

Another closely related model that also represents and learns topic correlations is hierarchical LDA (hLDA) (Blei et al., 2004). It is a variation of LDA that assumes a hierarchical structure among topics. Topics at higher levels are more general, such as stopwords, while the more specific words are organized into topics at lower levels. Note that each topic has a unique path from the root. To generate a document, hLDA samples a topic path from the hierarchy and then samples every word from those topics. Thus hLDA can well explain a document that discusses a mixture of *computer science*, *artificial intelligence* and *robotics*. However, for example, the document cannot cover both *robotics* and *natural language processing* under the more general topic *artificial intelligence*. This is because a document is sampled from only one topic path in the hierarchy. A nested Chinese restaurant process is used to model the topic hierarchy. It is different from the HDP-based prior in our model since it does not require shared mixture components among multiple DPs.

## 5 Conclusions and Future Work

In this paper, we have presented a nonparametric Bayesian prior for pachinko allocation based on a variant of the hierarchical Dirichlet process. While PAM could use arbitrary DAG structures to capture topic correlations, we focus on a four-level hierarchical structure in our experiments. Unlike a standard HDP mixture model, nonparametric PAM automatically discovers topic correlations from unstructured data as well as determining the numbers of topics at different levels.

As mentioned in Section 3, the topic structure discovered by nonparametric PAM is usually sparse. It allows us to pre-prune the unlikely sub-topics given a super-topic and dramatically reduce the sampling space. Therefore the training procedure will be more efficient, and we are interested in developing a scalable model that can be applied to very large datasets.

The four-level hierarchical structure is only a simple example of PAM. There are more complicated DAG structures that provide greater expressive power. Accordingly, it is more difficult to select the appropriate structures for different datasets and nonparametric approaches will be more appealing to such models. In our future work, we will explore in this direction to discover richer topic correlations.

**Acknowledgements**

This work was supported in part by the Center for Intelligent Information Retrieval and in part by the Defense Advanced Research Projects Agency (DARPA), through the Department of the Interior, NBC, Acquisition Services Division, under contract number NBCHD030010, and under contract number HR0011-06-C-0023 and in part by The Central Intelligence Agency, the National Security Agency and National Science Foundation under NSF grant number IIS-0326249. Any opinions, findings and conclusions or recommendations expressed in this material are those of the author(s) and do not necessarily reflect those of the sponsor.

## References

Blei, D., Griffiths, T., Jordan, M., & Tenenbaum, J. (2004). Hierarchical topic models and the nested Chinese restaurant process. In *Advances in neural information processing systems 16*.

Blei, D., & Lafferty, J. (2006). Correlated topic models. In *Advances in neural information processing systems 18*.

Blei, D., Ng, A., & Jordan, M. (2003). Latent Dirichlet allocation. *Journal of Machine Learning Research, 3*, 993–1022.

Daume-III, H., & Marcu, D. (2005). A Bayesian model for supervised clustering with the Dirichlet process prior. *Journal of Machine Learning Research 6*, 1551–1577.

Diggle, P., & Gratton, R. (1984). Monte Carlo methods of inference for implicit statistical models. *Journal of the Royal Statistical Society*.

Ferguson, T. (1973). A Bayesian analysis of some nonparametric problems. *Annals of Statistics, 1(2)*, 209–230.

Griffiths, T., & Steyvers, M. (2004). Finding scientific topics. *Proceedings of the National Academy of Sciences* (pp. 5228–5235).

Kim, S., Tadesse, M., & Vannucci, M. (2006). Variable selection in clustering via Dirichlet process mixture models. *Biometrika 93, 4*, 877–893.

Li, W., & McCallum, A. (2006). Pachinko allocation: DAG-structured mixture models of topic correlations. *International Conference on Machine Learning (ICML)*.

Sudderth, E., Torralba, A., Freeman, W., & Willsky, A. (2005). Describing visual scenes using transformed Dirichlet processes. *Advances in Neural Information Processing Systems 17*.

Teh, Y., Jordan, M., Beal, M., & Blei, D. (2005). Hierarchical Dirichlet processes. *Journal of the American Statistical Association*.

Xing, E., Sharan, R., & Jordan, M. (2004). Bayesian haplotype inference via the Dirichlet process. *International Conference on Machine Learning (ICML)*.